\documentclass[twocolumn,showpacs,preprintnumbers,amsmath,amssymb]{revtex4}

\input epsf
\newlength{\plotwidth}
\newcommand{\eff}{ {\rm{eff}} }

\usepackage{graphicx}

\begin{document}


\title{Flame Enhancement and Quenching in Fluid Flows}

\author {
Natalia Vladimirova$^\dagger$,
Peter Constantin$^\ddagger,$
Alexander Kiselev$^*,$
Oleg Ruchayskiy$^\dagger$
and
Leonid Ryzhik$^\ddagger$
}

\address {
$^\dagger$ASCI/Flash Center, The University of Chicago, Chicago, IL 60637}
\address{
$^\ddagger$Department of Mathematics, The University of Chicago, Chicago, IL
60637}
\address{$^*$Department of Mathematics, University of Wisconsin,
Madison, WI 53705}

\date{\today}

\begin{abstract}
  We perform direct numerical simulations (DNS) of an advected scalar
  field which diffuses and reacts according to a nonlinear reaction
  law. The objective is to study how the bulk burning rate of the
  reaction is affected by an imposed flow. In particular, we are
  interested in comparing the numerical results with recently
  predicted analytical upper and lower bounds. We focus on reaction
  enhancement and quenching phenomena for two classes of imposed model
  flows with different geometries: periodic shear flow and cellular
  flow. We are primarily interested in the fast advection regime. We
  find that the bulk burning rate $v$ in a shear flow satisfies $v
  \sim aU+b$ where $U$ is the typical flow velocity and $a$ is a
  constant depending on the relationship between the oscillation
  length scale of the flow and laminar front thickness.  For cellular
  flow, we obtain $v \sim U^{1/4}$. We also study flame extinction
  (quenching) for an ignition-type reaction law and compactly
  supported initial data for the scalar field. We find that in a shear
  flow the flame of the size $W$ can be typically quenched by a flow
  with amplitude $U \sim \alpha W$. The constant $\alpha$ depends on
  the geometry of the flow and tends to infinity if the flow profile
  has a plateau larger than a critical size. In a cellular flow, we
  find that the advection strength required for quenching is $U \sim
  W^4$ if the cell size is smaller than a critical value.
\end{abstract}


\maketitle


\section{Introduction}

Turbulent combustion in premixed flows is a widely studied topic in
both scientific and industrial settings (see e.g.
\cite{bray-peters94,Ronney,Peters}). The interest in the subject is due to an
important influence that advection can have on the reaction process:
both experimental \cite{bradley92,SRBY} and theoretical
\cite{CW,clavin-williams82,williams85,anand-pope87,yakhot88,kerstein88,
  poinsot-veynante-candel90,aldredge95,echekki-chen96,echekki-chen99}
work shows that the propagation speed of the flame can be
significantly altered by the fluid flow. Specifically, moderately
intense levels of turbulence have the tendency to accelerate the flame
speed $v$ beyond its laminar value $v_\circ$. The mechanism and the
extent of the flame acceleration depend on the particular regime of
burning \cite{borghi85}. The general reason for the enhancement is
that the fluid motion distorts the flame front, increasing the
reaction area. On the other hand, if the advection is too strong, it
can lead to the flame extinction. The critical strength of advection
which leads to quenching depends on the extent of the flame, strength
of reaction and diffusion, and properties of the flow.

At this stage, it is unreasonable to expect a complete analytical
theory describing the process of combustion in a fluid phase.
Indeed, detailed modelling of the phenomena involves solving a
reaction-diffusion system involving temperature (or energy) and
concentrations of reactants coupled with compressible
Navier-Stokes equations describing motion of the mixture
\cite{Peters,ZBLM}. Therefore, most of the studies in this field
which seek analytical conclusions use heuristic reasoning or
simplified models, which may approximately describe the system in
certain combustion regimes. Some of the combustion regimes are
relatively well-understood, such as the so-called flamelet regime,
where flame thickness is small compared to the fluid velocity
scales. The geometric optics approximation where the propagation
of the front is ruled by Huygens principle is often used as a
starting point in the analysis of this regime (see e.g.
\cite{Pel,kerstein-ashurst-williams88}).

Our goal here is to study one of the most widely used PDE models of
combustion, namely the scalar reaction-diffusion equation with passive
advection:
\begin{equation}
\frac{\partial T } {\partial t} + {\bf u} \cdot \nabla T = \kappa
\nabla^2 T + \frac{1}{\tau} R(T). \label{eqnT}
\end{equation}
Here $T$ is the normalized temperature, $0 \leq T \leq 1$, ${\bf u}$
is the fluid velocity, which we assume is incompressible, $\kappa$ is
thermal diffusivity, and $\tau$ is the typical reaction time.  In the
absence of fluid velocity Eq.~(\ref{eqnT}) admits flat propagation
front with laminar burning velocity of the order of $v_\circ \sim
\sqrt{\kappa/\tau}$ and characteristic thickness of the order of
$\delta \sim \sqrt{\kappa \tau}$. The model~(\ref{eqnT}) can be
derived from a more complete system under assumptions of constant
density and unity Lewis number (the ratio of material and temperature
diffusivity), as shown, for instance, in~\cite{CW}. The
equation~(\ref{eqnT}) has a more general applicability than the
geometrical optics approximation; moreover, as we will discuss below,
the geometrical optics limit can be obtained from~(\ref{eqnT}) in a
certain parameter range.

We will consider reaction rates $R(T)$ of two types, KPP (Kolmogorov,
Petrovskii, Piskunov) \cite{KPP,Fisher}, and ignition.  The KPP type
is characterized by the condition that the function $R(T)$ is positive and
convex on the interval $0<T<1$. This reaction type is used often in
problems on population dynamics (see e.g. \cite{AW1,Fi}), but is
relevant in combustion modelling, for example in some autocatalyctic
reactions \cite{HSS}.  A reaction term of ignition type is
characterized by the presence of critical ignition temperature, such
that the function $R(T)$ is identically zero below ignition
temperature. This type of reaction term is used widely to model
combustion processes (see e.g. \cite{VVV,ZBLM}), in particular
approximating the behavior of Arrhenius-type chemical reactions which
vanish rapidly as temperature approaches zero.

Our main goal is to gain insight into the question of how the
geometry and the amplitude of the fluid flow influence the
combustion process. Our study is partly motivated by recent
analytical work \cite{ABP,CKOR,CKR,HPS,KR} where rigorous bounds
on combustion enhancement and quenching are proved. We test the
sharpness of results in \cite{ABP,CKOR,CKR,HPS,KR}, and in
addition derive new predictions. We consider two classes of flows.
The first is shear flows, a representative of a wider class of
flows, called ''percolating'' in \cite{CKOR}, which have open
streamlines connecting distant regions of the fluid. The second
class is cellular flow, where the streamlines are closed and the
flow consists of isolated cells.  For each class of flows, we
study both flame enhancement and quenching.

For flame enhancement study, we consider initial temperature in
the form of the laminar front, with $T=1$ in the semi-infinite
region behind the front and $T=0$ in the the semi-infinite region
ahead of the front. Distorted by imposed flow, the flame front
propagates as a travelling wave with velocity higher than laminar.
The goal of the flame enhancement study is to obtain relations
between flame propagation speed $v$ and the properties of the
flow, especially for the large advection velocities.


In the case of quenching phenomena, we consider initial temperature to
be non-zero in a finite region. Since quenching cannot occur for the
KPP-type source term \cite{Roq,Roq-2}, we use the ignition-type
reaction term.  As shown by Kanel, there is a critical size $W_\circ$
of initial hot region below which the flame will be extinguished by
diffusion alone, e.g. with no advection, when the temperature drops
below the threshold and reaction ceases before the flame establishes a
steady travelling wave configuration \cite{Kanel}. When advection is
present, the fluid flow stretches the initially hot region so that it
can be quenched by diffusion; hot regions of the size much larger than
$W_\circ$ can be quenched in this manner. Our goal has been to
understand how the geometry and amplitude of the flow influence the
size of the band $W$ of the initial hot region that can be quenched.



\section{Numerical Setup and Method}

\begin{figure}[b]
  \centerline{\includegraphics[width=6.cm]{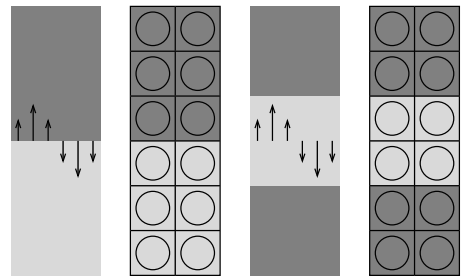}}
  \caption{Schematic representation of initial conditions and velocity
    field. Problem setup, from left to right: reaction enhancement in
    shear flow; reaction enhancement in cellular flow; quenching in
    shear flow; quenching in cellular flow. Dark tone corresponds
    to the cold fluid ($T=0$), light -- to the hot fluid ($T=1$).}
  \label{fgIC}
\end{figure}

The simulation is set in two space dimensions, in a vertical strip
of width $L$ with periodic boundary conditions in $x$ direction
(Fig.~\ref{fgIC}). In reaction enhancement studies, the initial
temperature was set to $T=1$ in the lower half of the domain and
to $T=0$ in the upper half of the domain. In the quenching studies
the initial temperature was set to $T=1$ in a horizontal band of
width $W$ in the center of the domain, and to $T=0$ elsewhere. The
interfaces between hot and cold fluid were smoothed to match
the laminar flame thickness.

We consider two types of flows, sinusoidal shear flow with amplitude
$U$ and wavelength $L$, perpendicular to the initial temperature front(s),
\begin{equation}
 {\bf u} = U \left( 0, \; \cos\frac{2\pi x}{L} \right),
 \label{eqShear}
\end{equation}
and cellular flow with amplitude $U$ and wavelength $L$,
\begin{equation}
{\bf u} = U \left(\sin \frac{2\pi x}{L} \cos\frac{2\pi y}{L}, \;
           - \cos \frac{2\pi x}{L} \sin \frac{2\pi y}{L}\right).
\label{eqCell}
\end{equation}
In the quenching simulations, the size of the cell $L/2$ was a fraction
of $W$, so that the initial band always contains integer number of
cells.

Most of the reaction enhancement computations were done using KPP
reaction rate \cite{KPP,Fisher} in the
advection-reaction-diffusion equation~(\ref{eqnT}),
\begin{equation}
  R(T)= \frac{1}{4} \, T(1-T),
  \label{eqKPP}
\end{equation}
with some of the simulations repeated with ignition type reaction,
\begin{equation}
  R(T)=\frac{T_\circ}{(1-T_\circ)^2} \, (1-T), \hspace{10mm} T>T_\circ,
  \label{eqIgnition}
\end{equation}
where $T_\circ$ represents threshold temperature, below which
$R(T)=0$.  In quenching studies we use ignition type reaction
(\ref{eqIgnition}) with threshold temperature $T_\circ=0.5$. Both
reaction rates (\ref{eqKPP}) and (\ref{eqIgnition}) were chosen to
exactly match, in the absence of advection, laminar burning velocity
$v_\circ = \sqrt{\kappa/\tau}$.
The corresponding laminar flame thickness is in both cases
of the order of $\delta = \sqrt{\kappa \tau}$; for KPP reaction it is
several times wider than for ignition.

Equation~(\ref{eqnT}) with reaction rates (\ref{eqKPP}) and
(\ref{eqIgnition}) has been solved using a fourth-order explicit
finite difference scheme in space and a third-order
Adams-Bashforth integration in time.  The grid size was chosen so
as to accurately represent the shear across the reacting region:
typically of the order of 12 zones across the flame interface for
thin fronts and at least 32 zones per period for thick fronts. The
computational domain extended a considerable distance upstream and
downstream from the burning front so that boundary effects were
negligible. In flame enhancement simulations the overall grid was
remapped following the propagation of the front, thereby allowing
for long integration periods --- of the order of 1000 reaction
times $\tau$.  We found that these long integrations were
necessary in order to reproduce correctly the asymptotic behavior
of the propagation speed in the case of strong advection.

As a measure of the reaction enhancement we use the bulk burning rate
\begin{eqnarray} \label{br}
  v(t) &=& \frac{1}{L}
 \int_0^L \int_{-\infty}^\infty \frac{\partial T(x,y,t)}{\partial t}
 \,dydx \nonumber \\
     &=& \frac{1}{\tau L} \int_0^L \int_{-\infty}^\infty R(T)\,dydx.
\end{eqnarray}
The second equality in (\ref{br}) can be justified by integration by
parts.  The bulk burning rate coincides with the front velocity in a
case where the solution is a travelling wave, but provides a more flexible
measure of combustion. Physically, $v(t)$ can be understood as the total
amount of reacted material or the total heat production. In all
simulations done for reaction enhancement, $v(t)$ approaches an asymptotic
value (for cellular flows one should average in time to arrive at this
value) and we denote this asymptotic value $v$.

We remark that in shear and cellular flows equation (\ref{eqnT})
admits travelling wave-type solutions, called pulsating fronts (see
e.g.  \cite{B,BLL,BH,Xin1,Xin2}). A rigorous stability theory for these
solutions exists but is not complete (see \cite{Xin3} for a recent
review). Except when quenching occurs, in our simulations we always
observed convergence of the solution to such waves, so the bounds on
$v$ also provide bounds for the propagation speeds of pulsating
fronts.

In the quenching studies, we measure the total amount of burned material
per wavelength,
\begin{equation}
w(t) = \frac{1}{L} \int_0^L \int_{-\infty}^{\;\infty} T(x,y) \;dy\;dx,
\label{eqTtotal}
\end{equation}
which can also be interpreted as the width of the non-perturbed
horizontal band with temperature $T=1$ surrounded by the fluid with
$T=0$.  This quantity is related to the bulk burning rate per
interface, $v(t)=\dot{w}(t)$.
Depending on initial conditions and flow parameters, $w(t)$ either
approaches an asymptotic value, e.g. $v(t)\rightarrow 0$, or increases
with constant rate, that is $v(t)\rightarrow v$ (for cellular flow, in
the time-averaged sense). In the first case we say that flame
quenches; the main objective of quenching simulations is to determine
under which conditions this happens.


\section{Shear flow: Reaction Enhancement}

We carried out simulations with the sinusoidal shear flow with
amplitude flow $U$, and wavelength, $L$, given by Eq.~(\ref{eqShear}).
In this section we are interested only in reaction enhancement
phenomena, and therefore for initial conditions we consider $T=1$ in
the semi-infinite domain $y<0$, and $T=0$ for $y>0$. We carried out
computations for both KPP and ignition type reactions, but did not
find significant differences in the qualitative behavior. The
numerical results presented in this section are obtained with KPP
reaction term (\ref{eqKPP}).

\begin{figure}[b]
\centerline{  \includegraphics[width=8.cm]{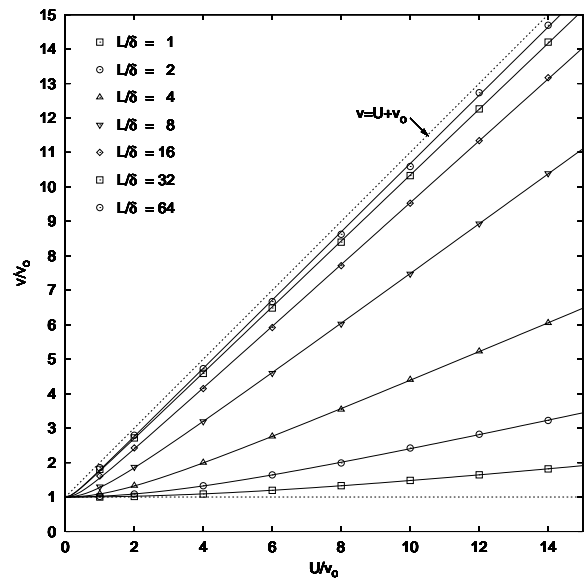}}
  \caption[]{
    Bulk burning rate (\ref{br}) as a function of the shear flow amplitude
    for different shear wavelengths.} 
  \label{fg1}
\end{figure}

Of special interest is the dependence of the effective propagation
rate $v$ on the velocity amplitude, $U$, and wavelength, $L$, which
defines the characteristic length scale of the flow (Fig.~\ref{fg1}).
For small amplitudes, $U \ll v_\circ$, our results are in agreement
with the quadratic law $v \sim v_\circ +cU^2$, which
goes back to Clavin and Williams \cite{CW} for turbulent
flow and has been recently proved rigorously for shear flows in
\cite{HPS}. We did not study this regime in detail since our main
interest is in the strong advection case.

For the amplitudes $U \gg v_\circ$ the results are in good
agreement with the linear law $v = aU + b,$ where the coefficients $a$
and $b \leq v_\circ$ depend on the geometry of the flow. In the
situation where the scale of the flow is much larger than the reaction
length scale, $L\gg\delta$, our data agrees with
$v=U+v_\circ.$ This law has been proposed in \cite{ABP} for shear
flows which vary slowly compared to the typical reaction length and
rigorously proved in \cite{CKR} under similar assumptions. For any
fluid flow, the regime $L\gg\delta$ is closely related to the
so-called geometrical optics combustion regime \cite{Ronney}, the
limit where reaction time and length scales approach zero. In the
framework of the equation (\ref{eqnT}) this corresponds to the limit
$\kappa, \tau \rightarrow 0$ while $\kappa/\tau$ remains constant.
Indeed, by rescaling equation (\ref{eqnT}) with a factor $L/\delta$ in
space and time,
we find that the bulk burning rate $v$ for the original equation
(\ref{eqnT}) is the same as for equation with modified diffusion and
reaction (\ref{effL})
\begin{equation}\label{effL}
T_t + {\bf u} \cdot \nabla T - \kappa \frac{\delta}{L} \Delta T =
\frac{L}{\delta \tau} R(T).
\end{equation}
 As $L$ grows, equation (\ref{effL})
approaches the geometrical optics limit.

\begin{figure}[tb]
  \centerline{\includegraphics[width=8.cm]{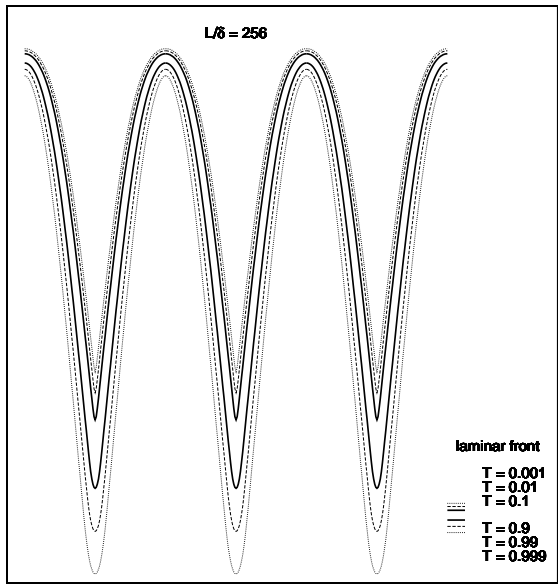}}
  \caption[]{
    Isotherms within the front in the geometrical optics limit.
    Here $L/\delta=256$, and $U/v_\circ=4$.  }
  \label{fg2'''}
\end{figure}

Quite often, front propagation in the thin front and fast reaction
limit is modelled by Hamilton-Jacobi type equations. One such model
 is the $G$-equation
\begin{equation}\label{Geq}
G_t + {\bf u} \cdot \nabla G = v_\circ |\nabla G|,
\end{equation}
where the front is defined by a constant level surface of the
scalar $G$ (see e.g. \cite{Pel}). The $G$-equation describes
propagation of the front according to Huygens principle; that is,
the front (i) is transported by fluid flow, and (ii) propagates
normal to itself with the speed $v_\circ$. The law $v=U+v_\circ$
can also be understood from the point of view of geometrical
optics since it is easily derived from the $G$-equation. Recently
Majda and Souganidis pointed out that the $G$-equation does not
provide the geometrical optics limit of the reaction diffusion
equation (\ref{effL}) in a precise sense \cite{MS1,EMS}. However
the rigorous bounds derived for the true effective equation
still give the same prediction for $v$ in the case of shear flow
\cite{MS1}.

In the situation where $L$ becomes comparable to $\delta$, the
coefficient proportionality $a$ between velocity amplitude and
flame propagation rate is no longer equal to unity. The rigorous
lower bound for $v$ from \cite{KR} takes form $v \geq C_1U
\frac{1}{1+C_2 n}$, where $n = 2\pi \delta/L.$ This bound is in
good qualitative agreement with an argument proposed by Abel,
Celani, Verni and Vulpiani \cite{ACVV1} based on the effective
diffusivity for the shear flow. It is well known that, if the
problem is considered on sufficiently large time and length
scales, the effect of the advection of passive diffusive scalar
can often be modelled by effective diffusivity \cite{AMa,AVer}.
The expression of effective diffusivity in a strong shear flow
goes back to Taylor \cite{Taylor}, 
$\kappa_\eff =\kappa +\frac{1}{2}\left(\frac{U}{v_\circ
      m}\right)^2\!\kappa$, where $m=2\pi \delta/l$ and $l$ is the
typical length scale of the flow. In the presence of reaction, we take
$l = {\rm min}(\delta, L),$ since the advection balances with reaction
instead of diffusion if $L>\delta$. This leads to the qualitative 
prediction 
$v \sim U$ if $L \gg \delta$ and $v \sim UL/\delta$ if $L \lesssim
\delta$. We obtained good although not perfect agreement with this
prediction. This is not surprising given the heuristic derivation of
the expression for the effective diffusivity and its possible
dependence on more subtle geometric properties of the flow.

\begin{figure}[tb]
  \centerline{\includegraphics[width=8.cm]{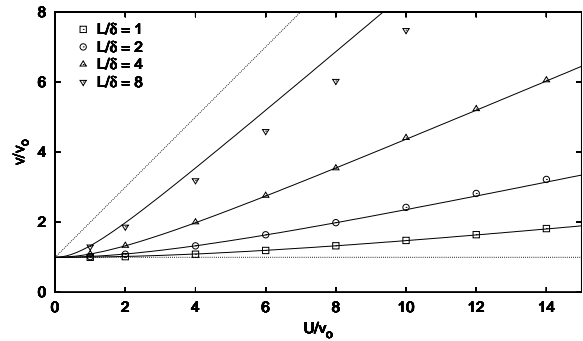}}
\caption[]{
  Bulk burning rate in high wavenumber sinusoidal shear flow as
  function of shear amplitude (points), compared with isotherm
  elongation (solid lines) given by Eq.~\ref{vsmall}.}
  \label{fg2'}
\end{figure}

\begin{figure}[tb]
\centerline{\includegraphics[width=8.cm]{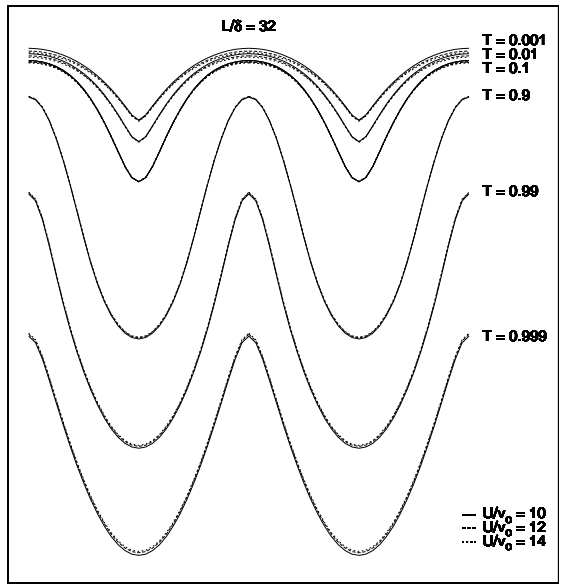}}
\caption[]{
  Isotherms within the front for cases with $L/\delta=32$ and
  $U/v_\circ=10, 12, 14$. The individual curves have been rescaled by
  a factor of $(U/v_\circ)^{-1}$ in the $y$ direction. }
\label{fg2''}
\end{figure}

Additional understanding of linear dependence $v(U)$ can be gained
from studying the relationship between the burning enhancement and
the structure of the front, in particular level sets of the
temperature. Assume that in the geometrical optics approximation
the front is given by the function $y=f(x)$; then for the
travelling wave obeying Huygens principle and propagating with
speed $v$, we have
\begin{equation*}
  v =u(x) + v_\circ\sqrt{1+(f')^2},
\end{equation*}
where $u(x)$ is profile of the shear, $u(x) = U \cos \frac{2\pi
  x}{L}$. In the case where $u$ is a mean zero flow, this leads to the
expression
\begin{equation}
   v = \frac{v_\circ}{L} \int_0^L \sqrt{1+(f')^2}\,dx.
  \label{vsmall}
\end{equation}
Thus, we obtain a well-known fact that the speed of propagation is
proportional to the area of the front which in geometrical optics
limit coincides with a level set of $T$ (see Fig. \ref{fg2'''} for a
picture of level sets in a situation close to geometrical optics). It
is interesting to test to what extent this relationship remains true
in the situations where geometrical optics regime is no longer valid,
for example when $L$ is comparable to $\delta$. We found that for
large $U$, there is still good agreement between the elongation factor
of the level sets of temperature and combustion enhancement (Fig.
\ref{fg2'}).  Moreover, we found that for large $U$ the temperature
distribution across the front scales with the shear amplitude (see
Fig.  \ref{fg2''}), providing another explanation of the linear
dependence of the bulk burning rate on the amplitude of the flow.
This scaling behavior can be understood in terms of the approximate
self-similarity of equation (\ref{eqnT}) with respect to the change of
variables $\tilde{y}=\frac{y/\delta}{U/v_\circ}$,
$\tilde{x}=x/\delta$, $\tilde{t}=t/\tau$, which gives
\begin{equation}
  \label{nsa}
  \frac{\partial T}{\partial \tilde{t}} +
  \cos \left(\frac{2\pi x}{L} \right) \frac{\partial T}{ \partial
  \tilde{y}} = \frac{\partial^2 T}{\partial \tilde{x}^2} + 
  \left( \frac{v_\circ}{U} \right)^2
\frac{\partial^2 T} {\partial \tilde{y}^2} + R(T).
\end{equation}
The only term which now depends on $U$ is the one proportional to
the second derivative in $\tilde{y},$ and it becomes negligible in
the limit of large $U$. Indeed, equation (\ref{nsa}) without this term 
is hypoelliptic, and so addition of the second derivative term does not 
constitute a singular perturbation. That leads to $U$-independent propagation
rate $\tilde{v} \equiv \frac{v}{U} = \tilde{v}(L)$ and to linear
proportionality $v \propto U$  for large $U/v_\circ$.

We conclude this section by a remark that understanding of the
combustion enhancement in a shear flow appears to be useful in some
situations where flows with different structure are involved. In
particular, in the reactive Boussinesq system the flow consists of
vortices moving along with the reaction front \cite{VR?}. However in
the frame moving with the front the effect of such vortical flow is
similar to the shear. The prediction for the reaction enhancement in
such system based on the results for the shear flows appears to be in
agreement with numerically observed behavior \cite{VR?}.


\section{Cellular flow: Reaction Enhancement}

Cellular flows have been studied by many authors (see e.g.
\cite{Chandra,DR}), since similar fluid motions appear in many
important applications; classical examples are the two-dimensional
rolls of the Rayleigh-B\'enard problem and Taylor vortices in Couette
flow.

For cellular flow simulations, we use a velocity field given by
Eq.~(\ref{eqCell}).  The flow is controlled by two parameters,
velocity amplitude $U$ and wavelength $L$.  As in simulations of
reaction enhancement by shear flow, we consider initial conditions
with $T=0$ in the upper half of the computational domain ($y<0$),
and $T=1$ in the lower half ($y>0$). Most of the results presented
in this section were obtained with KPP reaction term
(\ref{eqKPP}); the influence of the reaction type on the reaction
enhancement is discussed at the end of this section.

There are several regimes, which can be classified according to the
relations between the characteristic scales present in the problem.
There are three characteristic time scales: advection, $\tau_U = L/U$;
reaction, $\tau_R = \tau$; and diffusion, $\tau_D = L^2/\kappa$. In
this paper, we mostly studied two regimes: $\tau_U \ll \tau_R \ll
\tau_D$ and $\tau_R \ll \tau_U \ll \tau_D$. The first is the regime of
strong advection; in the second regime advection can be very strong as
well, but is compensated by large cell size $L,$ so that the reactive
time scale becomes the fastest in the problem.  If $\tau_D \ll
\tau_R$, or equivalently, $L \ll \delta$, we have diffusive or small
cell regime. The remaining regime corresponds to $\tau_R \ll \tau_D
\ll \tau_U,$ and therefore $U \ll v_\circ$.  Hence we have slow
advection; this situation is of less interest to us since $v \leq
U+v_\circ$ under very general conditions \cite{CKOR}, so the effect of
the advection on the propagation speed is minor.

\begin{figure}[t]
  \centerline{\includegraphics[width=6.cm]{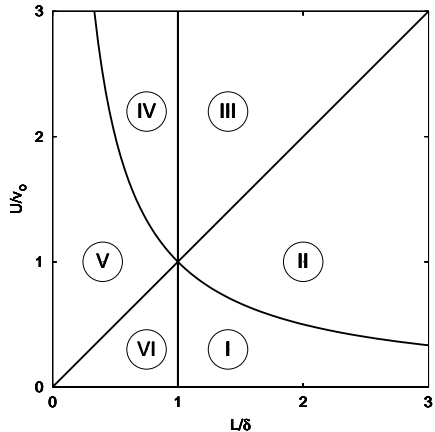}}
  \caption{ Burning regimes in cellular flow:
    (I)~$\tau_R<\tau_D<\tau_U$~--- slow advection geometrical optics;
    (II)~$\tau_R<\tau_U<\tau_D$~--- fast advection geometrical optics;
    (III)~$\tau_U<\tau_R<\tau_D$~--- fast advection with radial burning
                                  within cells;
    (IV)~$\tau_U<\tau_D<\tau_R$~--- fast advection with uniform burning
                                  within cells;
    (V)~$\tau_D<\tau_U<\tau_R$~--- limited advection in small cells;
    (VI)~$\tau_D<\tau_R<\tau_U$~--- slow advection in small cells.
    }
\end{figure}

\begin{figure}[t]
  \centerline{\includegraphics[width=8.cm]{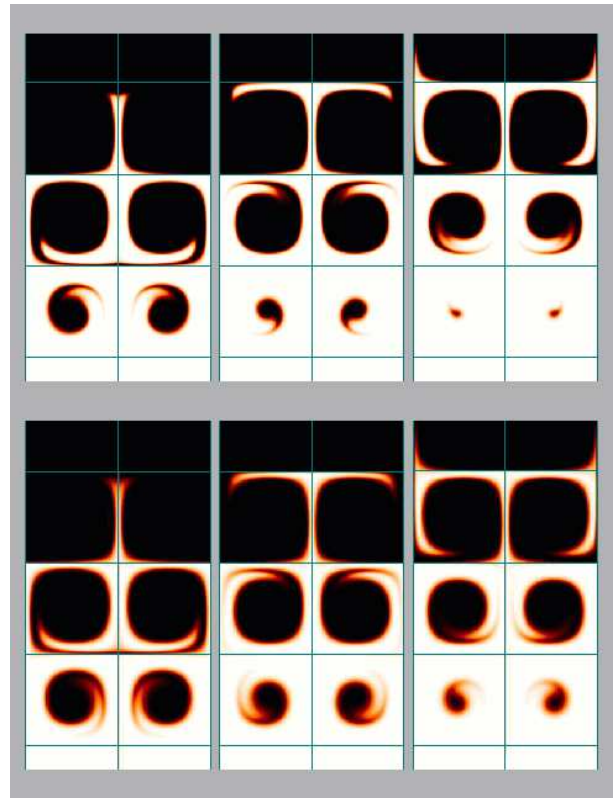}} \caption[]{
    Flame in cellular flow with amplitude $U/v_\circ=20$ and period
    $L/\delta=1024$ (upper row) and $L/\delta =512$ (lower row).
    Snapshots for the first case were taken with time interval
    $24\,\delta/v_\circ$ and for the second - with time interval
    $12\,\delta/v_\circ$.}
\label{fg4}
\end{figure}

\begin{figure}[t]
  \centerline{\includegraphics[width=4.cm]{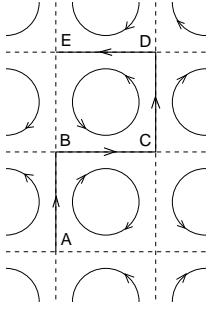}} \caption[]{
    Approximate path of the tip of the flame}
\label{fg3}
\end{figure}

\begin{figure}[b]
  \centerline{\includegraphics[width=8.cm]{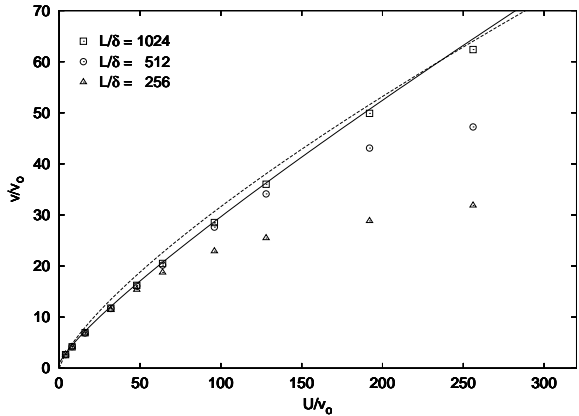}} \caption[]{
    Flame propagation velocity in thin flame regime
    as function of the cellular flow amplitude. Solid line is doubled
    Eq.(\ref{eqUlogU}), dashed line is $v/v_\circ = (U/v_\circ)^{3/4}$.}
\label{fg5}
\end{figure}

In the regime $\tau_R \ll \tau_U \ll \tau_D$ our simulations show
good agreement with geometrical optics models. In Figure~\ref{fg4}
we show a typical picture of flame in the regime close to
geometrical optics. The $G$-equation, (\ref{Geq}), which is closely 
related to geometrical optics regime, is invariant under simultaneous 
rescaling
of time and space by the same factor. This suggests that for
flames approaching the geometrical optics limit, one should
observe this similarity, and indeed we do as shown in
Fig.~\ref{fg4}. The earliest prediction of the front propagation
speed in a cellular flow for large $U$ within the geometrical
optics framework appears to be due to Shy, Ronney, Buckley and
Yakhot \cite{SRBY}. Using heuristic reasoning, they proposed that
$v \sim U/\log (U/v_\circ)$ if one considers front advancing
according to Huygens principle.  The same law is proposed in
\cite{ACVV2} based on more detailed analysis. To illustrate the origin
of this law,
let us sketch an argument providing the lower bound for $v.$ Let
us look at the propagation of the flame tip along the path ABCDE
on the Fig.~\ref{fg3}.  Assuming that at every point of ABCDE the
flame velocity is given by $u(x,y)+v_\circ,$ and integrating in
time, we obtain a lower bound
\begin{equation}
  \frac{v}{v_\circ} \geq \frac{\pi}{4}
  \frac{\sqrt{\left( \frac{U}{v_\circ} \right)^2
      -1}}{\log\left(\frac{U}{v_\circ}+\sqrt{\left(\frac{U}{v_\circ}\right)^2-1}
    \right)}.
  \label{eqUlogU}
\end{equation}
This lower bound, when doubled, gives a very good fit to our numerical
data and is represented by a solid line in Fig.~\ref{fg5}. As one can
see in Fig.~\ref{fg4}, the tip of the flame follows the path close to
ABCDE, but avoiding corners; this may account as a factor for the
difference in speed compared with the lower bound. Our results in the
geometrical optics regime agree with results of \cite{ACVV2}; however
we stress that our simulation was carried out for the
reaction-diffusion equation (\ref{eqnT}) using the same numerical
scheme in all regimes, while \cite{ACVV2} uses a different model in
geometrical optics limit.  We remark that in \cite{ACVV1,ACVV2} the
possibility of the regime, where $\tau_R \ll \tau_U$ and $v$ behaves
as $U^{3/4}$, was proposed. We could not definitively confirm existence
of such a regime due to the closeness of $(U/v_\circ)^{3/4}$ and
$(U/v_\circ)/ \log (U/v_\circ)$ curves in the range of tested
parameters (Fig. \ref{fg5}).

\begin{figure}[t]
  \centerline{\includegraphics[width=8.cm]{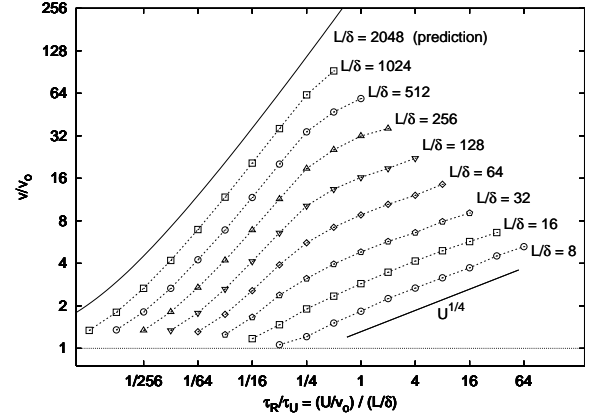}} \caption[]{
    Flame propagation velocity as function of the ratio of laminar
    burning time to the vortex turnover time. The $L/\delta=2048$
    prediction is based on doubled Eq.~(\ref{eqUlogU}) for
    geometrical optics.}
\label{fg6}
\end{figure}

\begin{figure}[b]
  \centerline{\includegraphics[width=8.cm]{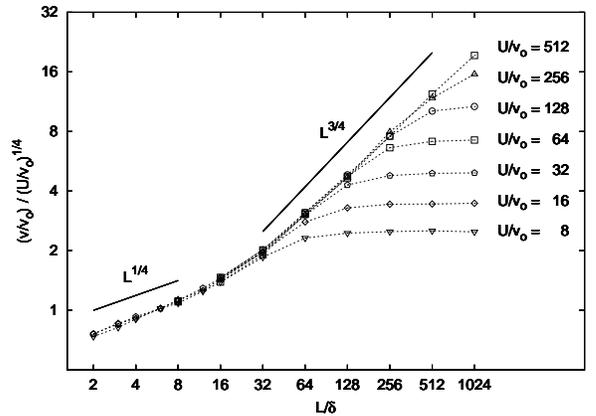}} \caption[]{
    Dependence of the flame propagation velocity on the size of the
    vortex.} \label{fg7}
\end{figure}

It should be emphasized that the geometrical optics regime
requires not only the thin front assumption, $L \gg \delta$, but
also fast reaction in comparison with advection, $\tau_R \ll
\tau_U$; in other words, velocity must be limited by $U \ll
(L/\delta)\,v_\circ$. When this restriction is broken we observe
significant decrease in flame propagation speed compared to the
geometrical optics prediction (Fig.~\ref{fg5}). Figure~\ref{fg6}
further illustrates this point, showing that on a logarithmic
scale, there is a marked change in the slope of $(v/v_\circ)$ as a
function of $U$ as $\tau_R/\tau_U$ increases. When $\tau_U$
exceeds $\tau_R$, we observe power-law, $v \sim U^{1/4}$, as
proposed by Audoly, Berestycki and Pomeau \cite{ABP}, and
confirmed in \cite{ACVV2} in a narrower range of parameters.  The
measurement of the slope $1/4$ was sufficiently precise to
distinguish it from the $v \sim U^{1/5}$ behavior, a lower bound
rigorously proved in \cite{KR}. The observed $v \sim U^{1/4}$
scaling extends to the limit of cell sizes small in comparison
with the laminar thickness, $L \ll \delta$. We remark that the
laminar front thickness for the KPP reaction is of the order of
$16\,\delta$ which is large compared to smallest cell size
$L/2=4\,\delta$ shown in Fig.~\ref{fg6}; limited data available
for $L/\delta=4,2$ (shown in Fig.~\ref{fg7}) also confirms the $v
\sim U^{1/4}$ scaling. In the very small cell regime, the $v \sim
U^{1/4}$ scaling was rigorously proven in \cite{HPS} using
homogenization approach.

\begin{figure}[t]
  \centerline{\includegraphics[width=8.cm]{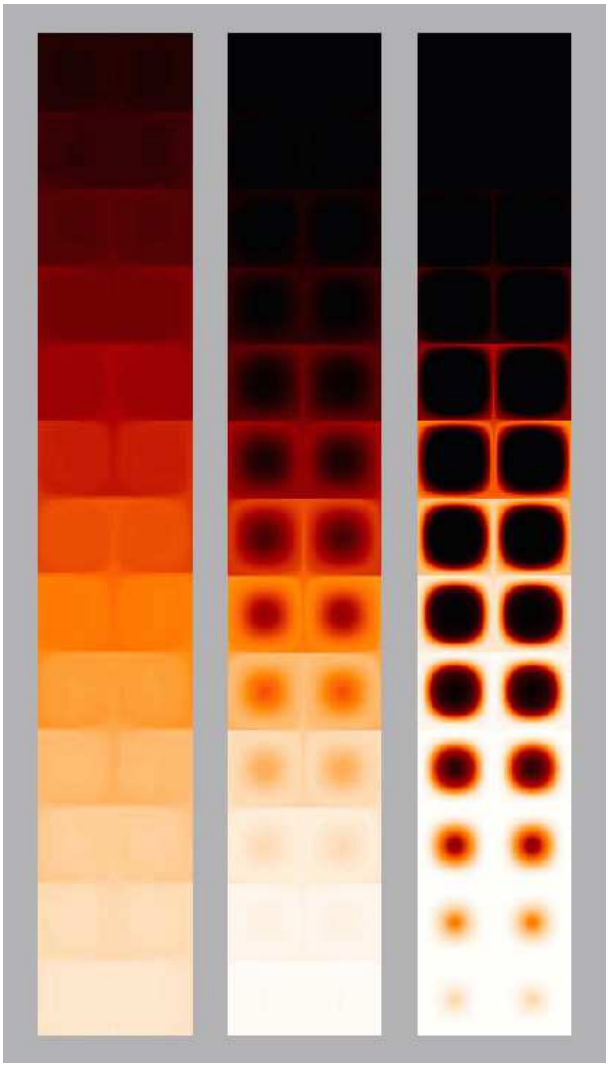}} \caption[]{
    Flame in a cellular flow with amplitude $U/v_\circ=100$ and period
    $L/\delta=16,64,256$ (left to right)} \label{fg8}
\end{figure}

\begin{figure*}[t]
  \centerline{\includegraphics[width=\textwidth]{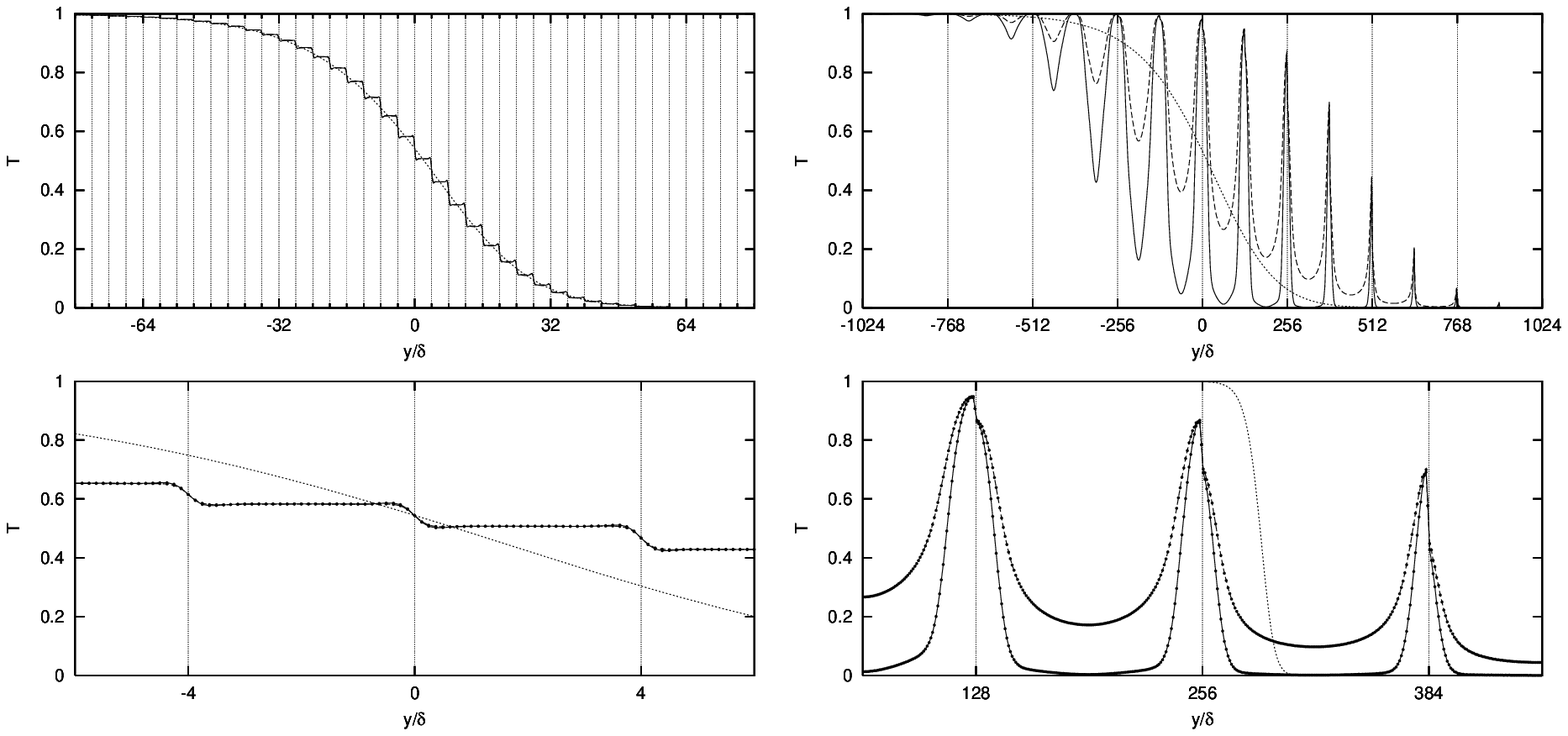}}
  \caption[]{
    Temperature profile at the middle of the cell (solid) and
    temperature averaged in $x$-direction (dashed) for $L/\delta=8,
    U/v_\circ=64$ (left) and for $L/\delta=256, U/v_\circ=128$
    (right).  Bottom plots are blow up versions of top plots. Dotted
    line in the top plots represents laminar front stretched by factor
    $v/v_\circ$ in $y$-direction, while in the bottom plots it
    represents non-modified laminar front.
    } \label{fg8'}
\end{figure*}

We also studied the dependence of $v$ on the cell size while
$U/v_\circ$ is fixed. Figure \ref{fg8} shows changes in the structure
of the flame with the increase of the cell size --- from a more
diffusive front to a front approaching geometrical optics behavior.
The flame propagation speed, normalized by $(U/v_\circ)^{1/4}$ factor,
is presented in Fig.~\ref{fg7}. As $L/\delta$ increases, we see the
transition from the $\tau_U \ll \tau_R$ regime, to the geometrical
optics, where flame propagation speed is independent of cell size. For
$\tau_U \ll \tau_R$ the data collapse to a single curve, suggesting
the power scaling with $L/\delta$, with power changing from $1/4$ for
small $L/\delta$ to $3/4$ for large $L/\delta$. The resulting scaling
can be summarized as,
\begin{eqnarray}
  v/v_\circ & \sim & (U/v_\circ)\, / \,\log(U/v_\circ),
  \hspace{3mm} \tau_R \ll \tau_U \ll \tau_D, \label{eqGop} \\
  v/v_\circ & \sim & (U/v_\circ)^{1/4}\,(L/\delta)^{3/4},
  \hspace{3mm} \tau_U \ll \tau_R \ll \tau_D. \label{eqLlg} \\
 v/v_\circ & \sim & (U/v_\circ)^{1/4}\,(L/\delta)^{1/4},
  \hspace{3mm} \tau_U \ll \tau_D \ll \tau_R, \label{eqLsm} 
\end{eqnarray}

To explain observed the flame propagation speed, let us consider a
model based on the effective diffusivity, proposed by Audoly,
Berestycki and Pomeau \cite{ABP}. When velocity is high, $\tau_U \ll
\tau_R$, the sharp temperature gradients appear in the narrow boundary
layer at cell borders (Fig.~\ref{fg8'}). The thickness of the boundary
layer, $h$, is determined by the balance of diffusion and advection,
$\kappa/h^2= U/L$, and is much smaller than $\delta$,
\begin{equation}\label{eqH}
h \sim \sqrt{\kappa L/U}.
\end{equation}
(This argument is a concise, naive version of the considerations
appearing in the derivation of the effective diffusivity in cellular
flow, \cite{Childress,YPP,Shraiman,RBDH}.)

A discrete diffusion equation modelling the original equation
(\ref{eqnT}) has been suggested in \cite{ACVV1,ACVV2}:
\begin{equation}\label{effcell}
  \frac{\partial \theta_n}{\partial t} =
  \frac{\kappa_\eff}{L^2}
  \left[\theta_{n-1}-2\theta_{n}+\theta_{n+1}\right] +
  \frac{1}{\tau} R(\theta_n).
\end{equation}
Here $\theta_n$ is the average
temperature in the $n$-th cell; and
$\kappa_\eff$ is effective diffusivity
\begin{equation}
  \kappa_\eff = \kappa \, L/h.
\end{equation}
This leads to the propagation rate,
\[
 v=\sqrt{\kappa_\eff/\tau} \sim v_\circ\sqrt{L/h},
\]
and, taking into account Eq.~(\ref{eqH}), to the scaling (\ref{eqLsm}).
We found that the prediction for speed given by (\ref{effcell}) agrees
with numerical simulations only if $L \lesssim \delta.$

For large cell sizes, $L \gg \delta$, the diffusive model
(\ref{effcell}) no longer accounts fully for the flame propagation
process. The main objection to the model is that (\ref{effcell})
assumes temperature is uniform inside the cell, and the
reaction term can be estimated at the average cell temperature.
However the numerically observed behavior demonstrates at first sharp
temperature gradients at the border of the cell, later evolving into
flame propagation inside of the cell roughly at the laminar flame
speed (see Fig.~\ref{fg8'} for the structure of the front inside
cells). Indeed, the cellular flow has no efficient mechanism for
mixing between the streamlines, and the diffusion time scale in that
direction is of the order $L^2/\kappa$ (practically not enhanced)
\cite{RY}. Therefore the combustion process inside the cell takes time
of the order of $L/v_\circ$, rather then $\delta/v_\circ$ (which
corresponds to substituting the average temperature into the reaction
term).

Here, we will modify the effective diffusivity model (\ref{effcell})
to account for slower burning inside large cells. As in the case of
small cells, the flame propagation from one cell to another is
enhanced because of the high temperature gradient in the boundary
layer with width given by expression (\ref{eqH}). The heat coming to
the cell through the cell boundary is distributed on the scale of
$\delta$ (as opposed to $L$, in the case of small $L$). We further
notice that due to the fast advection, the temperature is essentially
equal along the streamlines inside the cell (see Fig.~\ref{fg8}),
which allows the flame to propagate directly from one boundary layer
to another (Fig.~\ref{fg8'}).  That allows us to write the discrete
diffusion equation similar to (\ref{effcell}), but replacing averaging
in the cell by the averaging in the strip of width $\delta$ along the
cell border ($\delta$-layer),
\begin{equation}\label{efflayer}
  \frac{\partial \theta_n}{\partial t} =
  \frac{\kappa_\eff}{\delta^2}
  \left[\theta_{n-1}-2\theta_{n}+\theta_{n+1}\right] +
  \frac{1}{\tau} R(\theta_n).
\end{equation}
Here $\theta_n$ is the average of the temperature in the
$\delta$-layer of the $n$-th cell; and $\kappa_\eff$ is
effective diffusivity,
\begin{equation}
  \kappa_\eff = \kappa \, \delta/h.
   \label{effdiff}
\end{equation}
The temperature in a laminar front varies on the scale $\delta,$
and so estimating reaction in a $\delta$-layer by $R(\theta_n)$ is
justified. 
Equation (\ref{efflayer}) does not take into account the heat flux from 
the $\delta$-layer to the bulk of the cell. However, since $h \ll \delta,$
this heat flux does not enter the main balance.  
In other words, the front described by (\ref{efflayer})
propagates entirely in the
$\delta$-layers, with the rate
 $
  v_{\delta}=\sqrt{\kappa_\eff/\tau} \sim v_\circ\sqrt{\delta/h}.
$
Substituting $h$ from Eq.~(\ref{eqH}) we obtain,
\begin{equation}
   v_{\delta} \sim v_\circ \, ( U/v_\circ)^{1/4} (L/\delta)^{-1/4}.
   \label{eqVdelta}
\end{equation}

The time needed to ignite a new cell, e.g. to warm up a layer of
the size of the order $\delta$ in that cell, is equal to
$\tau_\delta = \delta / v_\delta.$ Once the width of a warmed up
layer reaches size of the order of $\delta$, reaction becomes
capable of sustaining the temperature.
Further propagation of the flame corresponds to the basically
laminar front movement inside the cell, and takes time $\tau_{\rm
cell} \sim L/v_\circ$.


\begin{figure}[b]
  \centerline{\includegraphics[width=8.cm]{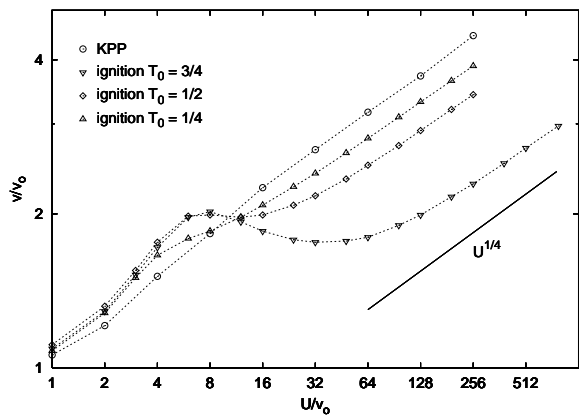}} \caption[]{
    Flame propagation velocity for different reactions ($L/\delta = 8$).}
\label{fgReac}
\end{figure}

The total bulk burning rate $v$ is of the order of $v_\circ$ times the
number of burning cells, which can be estimated as the ratio of the cell
burning time, $\tau_{\rm cell} \sim L/v_\circ$, to the time
needed to ignite a cell, $\tau_\delta  = \delta / v_\delta$.
Therefore the bulk burning rate is obtained by multiplying the
number of burning cells with $v_\circ,$ or essentially by
normalizing $v_\delta$ with a factor of $L/\delta$,
\begin{equation*}
  v \sim v_\circ \frac{\tau_{\rm cell}}{\tau_\delta} \sim
       \frac{L}{\delta}\, v_\delta \sim
       v_\circ \, ( U/v_\circ)^{1/4} (L/\delta)^{3/4},
\end{equation*}
which agrees with numerically observed scaling (\ref{eqLlg}).


Finally, we would like to mention the effect of the reaction rate.
Similar to the shear flow, we found that the reaction type does
not influence asymptotic scaling laws like (\ref{eqLsm}) or
(\ref{eqLlg}), although there is certainly a difference in the
constant factors. However, in cellular flows there is an
interesting phenomenon which is present for ignition-type but not
KPP reactions. The dependence $v(U)$ is always monotone increasing
in the KPP case, but it may exhibit a temporary reversal for
ignition-type reactions (Fig.~\ref{fgReac}). This effect has been
discovered by Kagan and Sivashinsky \cite{KS} and further studied
in \cite{KRS}. We found that this phenomenon is more pronounced
when the reaction threshold $T_\circ$ is closer to unity, in
agreement with the arguments of \cite{KRS}.


\begin{figure*}
  \centerline{\includegraphics[width=15cm]{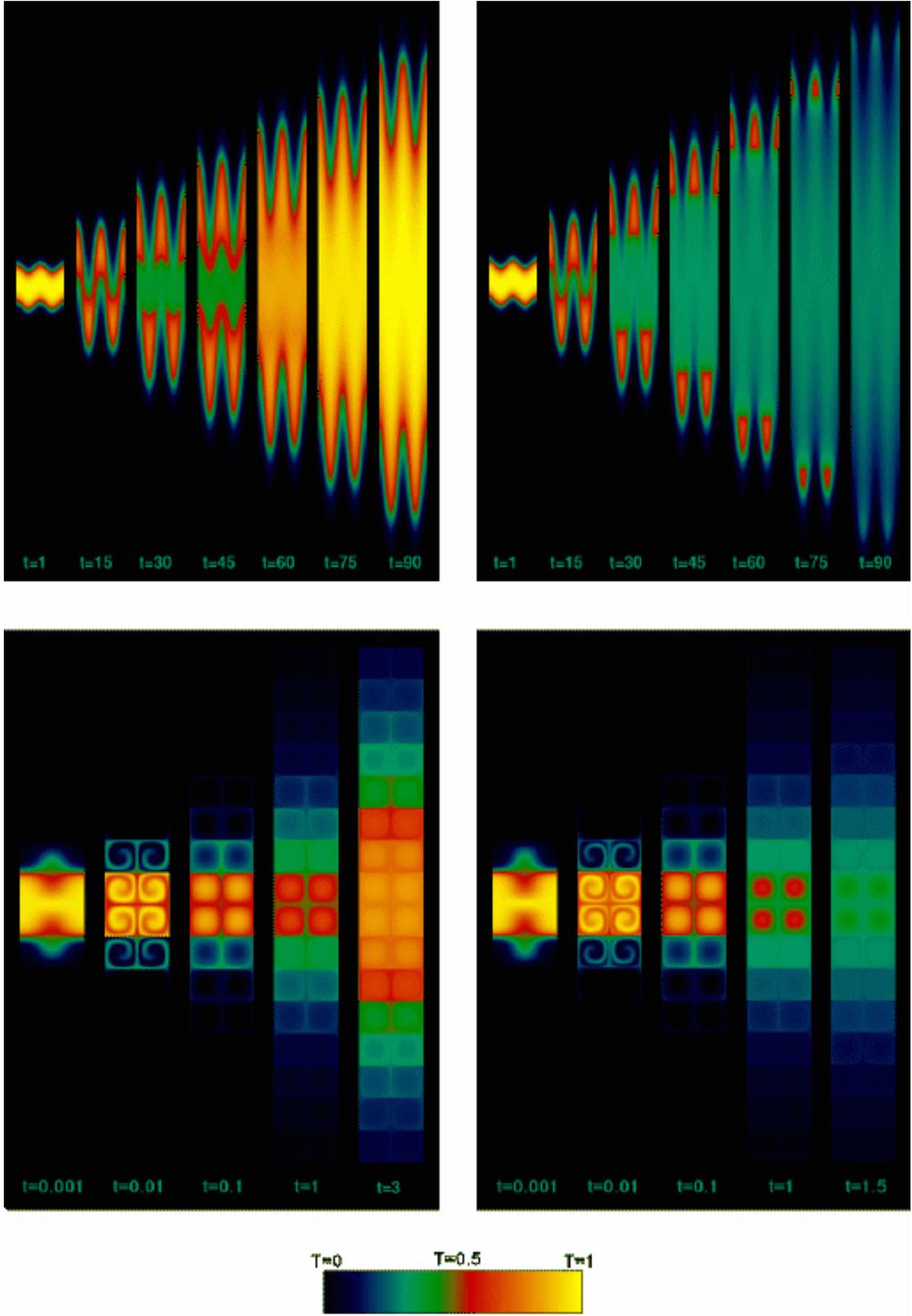}}
  \caption[]{Sequence of snapshot of temperature distribution 
    in the shear flow with $L/\delta=4$ (top) and in the cellular
    flow with $L/\delta=4$ (bottom). Initial condition was a hot band
    of the width $W/\delta=6$ and $W/\delta=4$ for shear and
    cellular flow simulations respectively. Velocities amplitudes are
    below critical on the left and above critical on the right (shear:
    $U/v_\circ=13$ and $U/v_\circ=14$; cellular: $U/v_\circ=600$
    and $U/v_\circ=800$). The time is given in units of $\tau$.}
    \label{fgl}
\end{figure*}

\section{Quenching}

%
%
%
%
%
In this section we address another effect that advection can have on
the combustion process --- quenching. We say that reaction is quenched
if the average temperature goes to zero uniformly with time.
Quenching occurs in the systems with ignition type reaction when,
due to diffusion and advection, temperature drops below the ignition
threshold everywhere and the integrated reaction rate becomes identically
zero.  

If the size of the region is small enough quenching can be caused by
diffusion alone, e.g. without advection, as shown by Kanel
\cite{Kanel}.  Kanel considered the one-dimensional reaction-diffusion
equation $T_t -\kappa T_{xx} = \tau^{-1} R(T)$. He found that there
exist two critical sizes $W_\circ \leq W^*_\circ$ such that if the
initial size of the hot region (where $T=1$) is smaller than
$W_\circ,$ reaction quenches, while if the initial size of the hot
region is greater than $W^*_\circ$, two fronts form and propagate in
opposite directions.  Two different critical sizes are likely an
artifact of the proof; in our simulations, we always found $W_\circ =
W^*_\circ \sim \delta$ and will refer to the single critical size
$W_\circ.$
In two and three dimensions, the presence of advection may
lead to stretching of the initial hot spot, thus making diffusion
more efficient at cooling, and consequently, at quenching.

In our numerical simulations we study quenching under the influence of
advection, in particular in shear and cellular flows. As in previous
sections, the prescribed flow velocities are defined by
Eq.~(\ref{eqShear}) for shear flow and by Eq.~(\ref{eqCell}) for
cellular flow. For all simulations we used the ignition-type reaction
(\ref{eqIgnition}) with threshold $T_\circ=1/2$, since quenching
cannot occur for the KPP-type source term \cite{Roq,Roq-2}. As the initial
conditions we use a horizontal band of width $W$ with temperature
above critical ($T=1$) within the band and with temperature below
critical ($T=0$) outside the band.

\begin{figure}[t]
  \centerline{\includegraphics[width=8.cm]{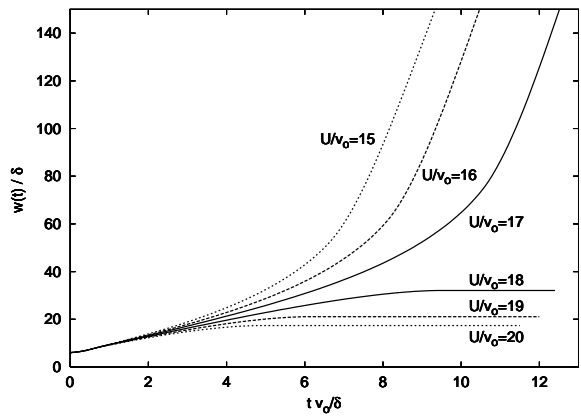}}
  \caption[]{ Temperature integrated over area $0<x<L$, $-\infty<y<\infty$
    per wavelength $L$ for different values of shear amplitude $U$,
    measured for $L/\delta=6$ and $W/\delta=6$.  } \label{fg11}
\end{figure}

The typical evolution of the system described above is shown in
Fig.~\ref{fgl}. We found that the temperature distribution in both for
shear and cellular flows evolves according to one of the two possible
scenarios, depending on the amplitude of the advection velocity.  For
lower advection velocities, after an initial transient period, the
system develops solution characterized by a wide, steadily growing
burned region between two wrinkled fronts propagating in opposite
directions. These fronts are exactly as described in the preceding
sections with regard to structure, speed, and dependence on the flow
properties~$U$ and~$L$. For higher advection velocities, the
temperature eventually drops below $T_\circ$ everywhere, after which
no burning occurs. We denote by $U_{\rm{cr}}$ the value of advection
velocity which triggers the system between these two scenarios
(further we refer to them as burning and quenching), and establish the
relation between $U_{\rm{cr}}$ and the initial conditions and
structure of the flow.

\begin{figure}[t]
  \centerline{\includegraphics[width=8.cm]{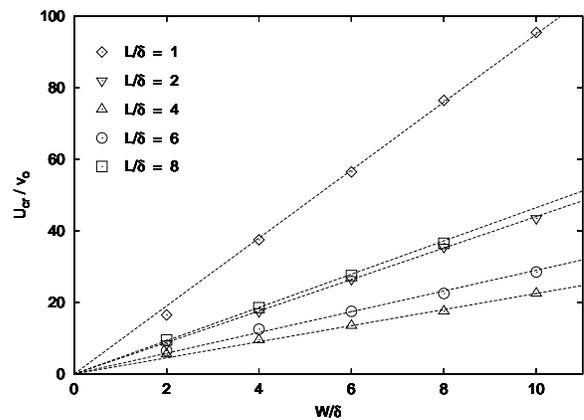}}
\caption[]{ The critical amplitude of shear flows with different
wavelengths as function of initial width of hot band.}
\label{fg12}
\end{figure}

\begin{figure}[b]
\centerline{\includegraphics[width=8.cm]{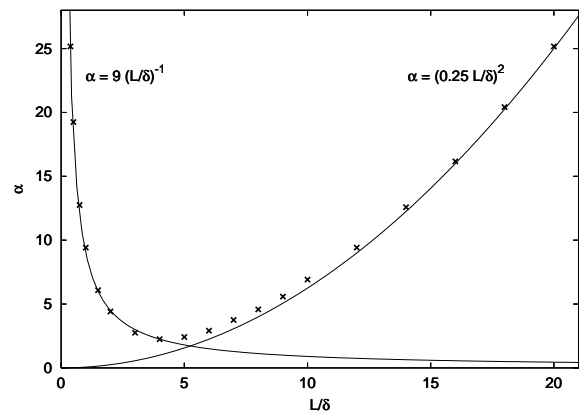}}
\caption[]{ Dependence of the factor $\alpha$ in Eq.~(\ref{eqUcrShear})
  on the wavelength of shear flow $L$. Measurements were taken at
  $W/\delta=6$} \label{fg13}
\end{figure}

We measure the critical value of velocity amplitude $U_{\rm{cr}}(L,W)$
using the following procedure. For each combination of the initial hot
band size $W$ and velocity period $L$, we execute a number of
simulations for different velocity amplitudes $U$. For each simulation
we measured the total amount of burned material per period $w(t)$,
defined by Eq.~(\ref{eqTtotal}), as a function of time (shown in
Fig.~\ref{fg11}). The burning systems (with higher velocities, where two front
are formed) are characterized in Fig.~\ref{fg11} by constant,
non-zero slopes, corresponding to constant reaction rate. These rates
are independent of initial conditions, and are equal to double the
burning rate $v(U,L)$, since there are two fronts. The quenched
systems are characterized by evolving to constant $w(t)$. Both
formation of steady fronts and quenched solutions requires some
transition time. 
 

The summary of results for a sinusoidal shear flow is given in
Fig.~\ref{fg12} and Fig.~\ref{fg13}.  We found that $U_{\rm{cr}}$ scales
linearly with $W$ (see Fig.~\ref{fg12}),
\begin{equation}
  U_{\rm{cr}} =\alpha \frac{W}{\tau},
  \label{eqUcrShear}
\end{equation}
as predicted in \cite{CKR}, with the coefficient $\alpha$ strongly
dependent on the wavelength of the advection velocity
(Fig.~\ref{fg13}).  Shear flow is most effective at quenching in the
intermediate range of wavelengths, namely when $L$ is of the order of
a few reaction lengths $\delta$. The quenching mechanism for small and
for large $L$ is different; one has to distinguish between the ability
of the flow to stretch the front over the larger scales and to make
the initial hot band uniformly thin.

For small $L$, the rapid spatial variation of the flow velocity is
well approximated by effective diffusion. The effective
diffusivity for strong shear flow scales as $\kappa_\eff \sim U^2
l^2/\kappa$, where $l={\rm min}(\delta,L)$. The characteristic
length scale for reaction with this renormalized diffusion behaves
as $ l_\eff \sim \sqrt{\kappa_\eff \tau}$ and scales linearly with
$U$. Then $l_\eff \sim W$ gives $U_{\rm{cr}} \sim
\frac{\delta}{L}W\tau^{-1}$ for small $L$. 
We remark that in the limit of small $L,$ the
effective diffusivity argument can be justified by a rigorous
homogenization procedure \cite{HPS}.

\begin{figure}[t]
  \centerline{\includegraphics[width=8.cm]{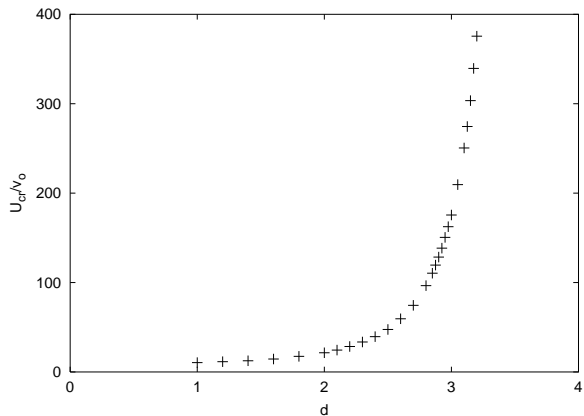}}
  \caption[]{ Growth in the value of $U_{\rm{cr}}$ for a fixed $W$ as the
    size of flat plateau increases. } \label{fg9}
\end{figure}

For large $L$, the nature of quenching is related to the appearance of
the (almost) constant regions in the velocity profile. We observe the
behavior $U_{\rm{cr}} \sim L^2$ for large $L,$ which can be explained
in the following way. In order for quenching to occur, the shear flow
should stretch the initial hot region thinner than Kanel's critical
length (of the order $\delta$) in time less than the reaction time
$\tau$, so that reaction does not have time to compensate cooling by
advection. The stretching is least efficient near the tip of the
velocity profile. At the tip of a sinusoidal profile, the difference
between flow velocities at two points separated by a distance $\delta$
is $U (\delta/L)^2$.  Therefore we obtain a sufficient condition for
quenching $ U_{\rm{cr}}\left( \frac{\delta}{L} \right)^2 \tau \sim W$,
which leads to
\[U_{\rm{cr}} \sim \tau^{-1} W \left( \frac{L}{\delta} \right)^2.\]

We also examined a degenerate case of shear flows with a plateau
in the velocity profile. For such flows it has been shown in
\cite{CKR} that quenching does not happen as soon as the size of
plateau is larger than certain critical size of the order $\delta$
and the size of initial bandwidth $W$ exceeds $W_\circ.$ As
expected, we found that $U_{\rm{cr}}$ diverged to infinity as the size
of the plateau approached a critical value (Fig.~\ref{fg9}), in
agreement with results of \cite{CKR}. This phenomenon can be
understood in terms of reaction and diffusion alone: in the region
where the profile of the velocity is flat, there is no stretching
of the initial hot band.  If the size of the hot band is roughly
larger than Kanel's critical size $W_\circ$, then reaction can
compete with diffusion, there will be no quenching, and eventually
propagating fronts will form.

\begin{figure}[t]
\centerline{\includegraphics[width=8.cm]{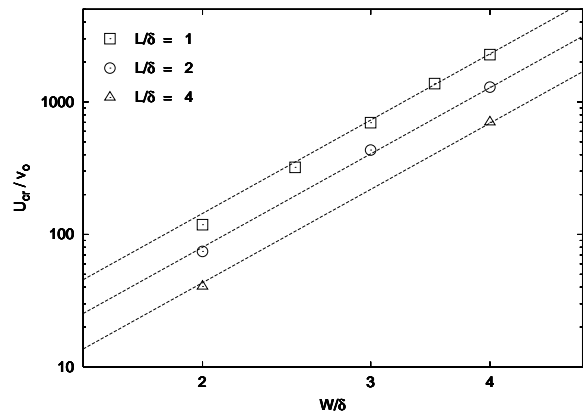}}
\caption[]{ The value of $U_{\rm{cr}}$ for cellular flows with
different periods.} \label{fg14}
\end{figure}

Quenching in cellular flow requires significantly higher advection
amplitudes. For relatively small cell sizes $L\lesssim\delta$ where
quenching is possible, we find that the critical velocity
$U_{\rm{cr}}$ satisfies $U_{\rm{cr}} \sim W^4$ (see Figure
\ref{fg14}).  Notice that this correlates with the dependence $v \sim
U^{1/4}$ for the speed of advection-enhanced flame. This correlation
is not coincidental, and can be explained as follows. The speed up law
indicates that the size of the region where reaction happens in a
stabilized combustion regime scales as
$(U/v_0)^{1/4}(L/\delta)^{1/4}\delta$ for large $U.$ If the width of
the initial band of hot material is of the order smaller than
$(U/v_\circ)^{1/4}(L/\delta)^{1/4}\delta$, advection carries away the
energy of the hot material over the larger region faster than reaction
is able to compensate the falling temperature. This leads to
quenching.

We remark that quenching is impossible if the cell size is
sufficiently large, $L \gtrsim \delta$, and $W \gtrsim \delta$. The
reason is similar to the flat plateau effect in the shear flow. Fluid
advection does not provide mixing inside cells in the direction
perpendicular to the streamlines, and thus if the cell is large enough
reaction can sustain itself against diffusion. This result has been
proved in \cite{CKR2}.

\section{Conclusions}

We carried DNS calculations of an advected scalar which reacts
according to a nonlinear reaction law. We studied combustion
enhancement and quenching phenomena in two typical classes of flows,
shear and cellular. In a shear flow, we find linear dependence
$v=aU+b$ of the combustion speed $v$ on the amplitude of the flow $U$
in the strong flow regime. The factor $a$ depends on the relationship
between the period of the flow $L$ and typical reaction length scale
$\delta,$ is equal to $1$ if $L \gg \delta$ and tends to zero if
$L/\delta \rightarrow 0.$ The observed behavior is in agreement with
recent rigorous \cite{CKOR} and numerical \cite{ACVV1} results. In
a cellular flow we studied primarily two regimes characterized by the
relationships $\tau_D < \tau_U < \tau_R$ and $\tau_D < \tau_R <
\tau_U$ between diffusion, reaction and advection time scales. We
found that combustion speed in the first regime is close to
predictions of models based on geometrical optics limit, $v \sim
U/\log(U/v_\circ).$ In the second regime where large $U$ dominates, we
found $v \sim v_\circ (U/v_\circ)^{1/4} (L/\delta)^{3/4}.$ This agrees
with the prediction of the effective diffusion model
\cite{ABP,ACVV1,ACVV2} in terms of the power of $U$ but has different
dependence the cell size $L.$ We proposed an explanation of the
observed behavior with a modified effective diffusion model where
enhanced diffusivity is concentrated in the boundary layers.

As opposed to combustion enhancement, quenching may happen if the
reaction term is of ignition type and initial temperature is higher
than critical in a finite region. If the shear flow velocity profile
does not have a plateau of sufficiently large size, or the
size of the cells in cellular flow is not too large, then for any
initial hot band size $W$ there exists $U_{\rm{cr}}(W)$ such that for
$U>U_{\rm{cr}}$ quenching takes place. If $U<U_{\rm{cr}}$, two fronts
form and propagate with the speed of the developed
advection-enhanced front. In the case of shear flow, $U_{\rm{cr}}$
depends linearly on $W$ with a factor $\alpha(L)/\tau$. Quenching is most
efficient for the flows with $L$ on the order of a few typical
reaction lengths $\delta$. For the cellular flows, $U_{\rm{cr}}$
scales as $W^4$. The results are in good agreement with theoretical
arguments.

\section{Acknowledgements}

 This research is supported in part by the
ASCI Flash center at the University of Chicago under DOE contract
B341495. PC was supported partially by NSF DMS-0202531. AK has
been supported by NSF grants DMS-0102554 and DMS-0129470 and
Alfred P. Sloan Fellowship. LR was supported partially by NSF
grant DMS-0203537 and by an Alfred P. Sloan Research Fellowship.

\end{document}